\documentclass[twocolumn,showpacs,nofootinbib,floatfix]{revtex4}
\usepackage{amssymb,dcolumn}
\usepackage[dvips]{graphics}
\usepackage{epstopdf}
\def\imo{i}

\def\noi{\noindent}

\def\nqq{\hspace*{-2em}}

\def\cm{\hspace*{1cm}}
\def\inch{\hspace*{1in}}

                         \def\d{\partial}
\def\lal{&&\nqq {}}                    
\def\eq{Eq.\,}                         
\def\eqs{Eqs.\,}                       
\def\beq{\begin{equation}}             
\def\eeq{\end{equation}}               
\def\bear{\begin{eqnarray}}            
\def\bearr{\begin{eqnarray} \lal}      
\def\ear{\end{eqnarray}}               
\def\earn{\nonumber \end{eqnarray}}    \def\kappa{\varkappa}

\def\nnn{\nonumber\\ \lal }            
\def\nnnv{\nonumber\\[5pt] \lal }      
                    \def\const{{\rm const}}
\def\yyy{\\[5pt] \lal }                
                     \def\ep{\epsilon}
               \def\then{\ \Longrightarrow \ }

\def\re{\mathop{\rm Re}\nolimits}
\def\im{\mathop{\rm Im}\nolimits}

\def\Jl#1#2{#1 {\bf #2},\ }

\def\ApJ#1 {\Jl{Astroph. J.}{#1}}
\def\CQG#1 {\Jl{Class. Quantum Grav.}{#1}}
\def\DAN#1 {\Jl{Dokl. AN SSSR}{#1}}
\def\GC#1 {\Jl{Grav. Cosmol.}{#1}}
\def\GRG#1 {\Jl{Gen. Rel. Grav.}{#1}}
\def\JETF#1 {\Jl{Zh. Eksp. Teor. Fiz.}{#1}}
\def\JETP#1 {\Jl{Sov. Phys. JETP}{#1}}
\def\JHEP#1 {\Jl{JHEP}{#1}}
\def\JMP#1 {\Jl{J. Math. Phys.}{#1}}
\def\NPB#1 {\Jl{Nucl. Phys. B}{#1}}
\def\NP#1 {\Jl{Nucl. Phys.}{#1}}
\def\PLA#1 {\Jl{Phys. Lett. A}{#1}}
\def\PLB#1 {\Jl{Phys. Lett. B}{#1}}
\def\PRD#1 {\Jl{Phys. Rev. D}{#1}}
\def\PRL#1 {\Jl{Phys. Rev. Lett.}{#1}}

\def\Veff{V_{\rm eff}}

\def\mn{_{\mu\nu}} 		\def\GR{general relativity}
 		
  		\def\dS{de Sitter}

\def\sph{spherically symmetric}
\def\ssph{static, spherically symmetric}
\def\bh{black hole}
\def\bhs{black holes}
\def\wh{wormhole}
\def\whs{wormholes}
\def\asflat{asymptotically flat}

\def\RN{Reissner-Nordstr\"om}
\def\KS{Kantowski-Sachs}
\def\Schr{Schr\"o\-din\-ger}

\begin{document}

\title{Instabilities of wormholes and regular black holes supported by a
	phantom scalar field}

\author{K. A. Bronnikov}
\email{kb20@yandex.ru}
\affiliation{Center for Gravitation and Fundamental Metrology, VNIIMS,
	Ozyornaya 46, Moscow 119361, Russia;\\
	Institute of Gravitation and Cosmology,
     	PFUR, ul. Miklukho-Maklaya 6, Moscow 117198, Russia}

\author{R. A. Konoplya}
\email{konoplya_roma@yahoo.com}
\affiliation{DAMTP, Centre for Mathematical Sciences, University of
	Cambridge, Wilberforce Road, Cambridge CB3 0WA, UK;\\
Centro de Estudios Cient\'{i}ficos (CECS), Casilla 1469, Valdivia, Chile.}

\author{A. Zhidenko}
\email{olexandr.zhydenko@ufabc.edu.br}
\affiliation{Centro de Matem\'atica, Computa\c{c}\~ao e Cogni\c{c}\~ao,
	Universidade Federal do ABC (UFABC),\\
	Rua Santa Ad\'elia, 166, 09210-170, Santo Andr\'e, SP, Brazil}


\begin{abstract}
  We test the stability of various wormholes and black holes supported by a
  scalar field with a negative kinetic term. The general axial perturbations
  and the monopole type of polar perturbations are considered in the linear
  approximation. Two classes of objects are considered: (i) \whs\ with flat
  asymptotic behavior at one end and AdS on the other (M-AdS \whs) and (ii)
  regular black holes with asymptotically de Sitter expansion far beyond the
  horizon (the so-called black universes).
  A difficulty in such stability studies is that the effective potential for
  perturbations forms an infinite wall at throats, if any. Its
  regularization is in general possible only by numerical methods, and such
  a method is suggested in a general form and used in the present paper.
  As a result, we have shown that all configurations under study are
  unstable under \sph\ perturbations, except for a special class of black
  universes where the event horizon coincides with the minimum of the area
  function. For this stable family, the frequencies of quasinormal modes
  of axial perturbations are calculated.
\end{abstract}

\maketitle

\bigskip
\section{Introduction}

  Modern observations \cite{Observation} indicate that the Universe is
  expanding with acceleration. The most favored explanation of this
  acceleration is nowadays that the Universe is dominated (to about 70\,\%)
  by some unknown form of energy density with large negative pressure,
  termed dark energy (DE), while the remaining 30\,\% consisting of baryonic
  and nonbaryonic visible and dark matter. It is often admitted that DE can
  be modeled by a self-interacting scalar field with a potential. Such a
  field acts as a negative pressure source; it is called quintessence if its
  pressure to density ratio $p/\rho = w >-1$ and a phantom field if $w < -1$
  while $w=-1$ corresponds to a cosmological constant $\Lambda$. Since
  observations admit a range of $w$ including $w=-1$, all sorts of models are
  under study.

  One should note that values $w < -1$ seem to be not only admissible but
  even preferable for describing an increasing acceleration, as follows from
  the most recent estimates: $w = -1.10 \pm 0.14$ ($1\,\sigma$)
  \cite{komatsu} (according to the 7-year WMAP data) and $w =
  -1.069^{+0.091}_{-0.092}$ \cite{sullivan} (mainly from data on type Ia
  supernovae from the SNLS3 sample). In this connection, cosmological models
  with phantom scalar fields, i.e., those with a negative kinetic term, have
  gained considerable attention in the recent years \cite{phantom}.

  If such fields can play an important role in cosmology, it is natural to
  expect that they manifest themselves in local phenomena, for instance, in
  the existence and properties of black holes and wormholes
  \cite{phantom-holes}. Quite a number of scalar field configurations of this
  kind have been described in the literature, see, e.g., examples of \bhs\
  with scalar fields (the so-called scalar hair) in \cite{sca-bh} and \whs\
  supported by scalar fields in \cite{h_ellis, kb73, pha1} and references
  therein. Thus, in \cite{pha1} it was shown that, in addition to \whs, a
  phantom scalar can support a regular \bh\ where a possible explorer, after
  crossing the event horizon, gets into an expanding universe instead of a
  singularity. Thus such hypothetic configurations combine the properties of
  a wormhole (absence of a center, a regular minimum of the area function)
  and a \bh\ (a Killing horizon separating R and T regions). Moreover, the
  Kantowski-Sachs cosmology that occurs in the T region is asymptotically
  isotropic and approaches a de Sitter regime of expansion, which makes such
  models potentially viable as models of our accelerating Universe. Such
  configurations, termed {\it black universes,\/} were later shown to exist
  with other scalar field sources exhibiting a phantom behavior such as
  k-essence \cite{pha4} and some brane world models \cite{Don10} (in the
  latter case, even without a phantom field).

  Both wormholes and black universes have been shown to exist as well in
  models where a scalar field exhibits phantom properties only in a
  strong-field region while in the weak-field region it obeys the canonical
  field equation (the so-called {\it trapped-ghost models\/})
  \cite{br-sush10,br-don11}

  To see whether or not such solutions can lead to viable models of black
  holes and wormholes, one needs to check their stability against various
  perturbations. Previously, gravitational stability as well as passage of
  radiation through wormholes supported by a phantom scalar field were
  considered, in particular, in \cite{Gonzalez:2008wd, Shinkai:2002gv,
  Gonzalez:2008xk, Doroshkevich:2008xm, Picon}, with a special emphasis on
  massless wormholes (see also references therein).

  In the present paper, we consider the linear stability of various \ssph\
  solutions to the field equations of general relativity with minimally
  coupled scalar fields, describing compact objects of interest such as
  asymptotically flat and AdS wormholes (M-AdS wormholes, for short,
  where M stands for ``Minkowski'') and black universes (in other words,
  M-dS regular black holes), and use as examples solutions obtained in
  \cite{pha1}. We show that M-AdS wormholes are unstable in the whole range
  of the parameters while among black universes there is a stable subfamily
  which corresponds to the event horizon located precisely at the minimum
  of the area function.

  The particular solutions whose stability is studied here are certainly
  not general. A more comprehensive study is prevented by the fact that a
  sufficiently general solution describing self-gravitating scalar fields
  with nonzero potentials is unknown, therefore it seems to be a natural
  decision to study the properties of known special solutions. On the other
  hand, in \cite{pha1} all possible regular \ssph\ solutions to the field
  equations were classified for phantom minimally coupled scalar fields with
  arbitrary potentials. One can see that the solution studied here, being
  certainly special, still represents a very simple but quite typical
  example reproducing the generic features of such configurations with flat,
  dS and AdS asymptotics; its another advantage is that it reproduces as a
  cpecial case the well-known Ellis wormhole, for which the stability
  results are already known \cite{Gonzalez:2008wd, Gonzalez:2008xk,
  Shinkai:2002gv}.

  The paper is organized as follows. Sec.\,II presents the backgrounds to be
  considered. Sec.\,III develops a general formalism for axial gravitational
  and Maxwell field perturbations in a static, spherically symmetric
  background. Sec.\,IV is devoted to polar spherically symmetric
  perturbations. Sec.\,V discusses the stability of the black universes and
  wormholes under consideration and analyzes the quasinormal radiation
  spectrum for the cases where the background configuration is linearly
  stable. In addition, we there develop a numerical tool for reducing
  the wavelike equation with a singular potential to the one with a regular
  potential. In Sec.\,VI we summarize the results and mention some open
  problems.

\section{Static background configurations}

  Let us consider Lagrangians of the form
\beq						\label{Lagrangian}
	L = \sqrt{-g} (R + \ep g^{\alpha \beta} \phi_{;\alpha}\phi_{;\beta}
	 - 2 V(\phi) - F_{\mu \nu} F^{\mu \nu}),
\eeq
  which includes a scalar field, in general, with some potential $V(\phi)$,
  and an electromagnetic field $F\mn$; $\ep = \pm 1$ distinguishes normal,
  canonical scalar fields ($\ep = +1$) and phantom fields ($\ep =-1$). In
  what follows, we present a perturbation analysis for static, spherically
  symmetric solutions for this general type of Lagrangian and then study the
  stability of some particular (electrically neutral) solutions.

  The general \ssph\ metric can be written in the form
\beq							   \label{metric}
	ds^2  = A(r) dt^2  - A(r)^{-1} dr^2  - R(r)^2 d\Omega^2,
\eeq
  where $d\Omega^2 = d\theta^2 + \sin^2 \theta d\varphi^2$ is the linear
  element on a unit sphere.

  We will consider the following static background \cite{pha1}:
\bearr
	R(r) = (r^2 + b^2)^{1/2}, \qquad b = \const > 0,
\nnnv                                                      \label{sol-pha1}
	A(r) = (r^2 + b^2)
\nnn\
	\times   \left[\frac{c}{b^2} + \frac{1}{b^2 + r^2}
    		+ \frac{3m}{b^3}\left(\frac{b r}{b^2 + r^2}
    			+ \tan^{-1} \frac{r}{b}\right) \right],
\nnn
	\phi = \sqrt{2} \tan^{-1} (r/b).
\ear
  It is a solution to the Einstein-scalar equations that follow from
  (\ref{Lagrangian}) with $F\mn \equiv 0$ and the potential
\bearr                                                     \label{V-pha1}
 	V(\phi) = -\frac{c}{b^2} \left[3 - 2 \cos^2
 		\left(\frac{\phi}{\sqrt{2}}\right)\right]
\nnn
	- \frac{3m}{b^3}\left[3 \sin \frac{\phi}{\sqrt{2}}
	\cos \frac{\phi}{\sqrt{2}}  + \frac{\phi}{\sqrt{2}}
		\left(3 - 2 \cos^2 \frac{\phi}{\sqrt{2}}\right) \right].
\ear

  The solution behavior is controlled by the scale $b$ and two integration
  constants: $c$ that moves the curve $B(r) \equiv A/R^2$ up and down, and
  $m$ showing the position of the maximum of $B(r)$. Both $R(r)$ and $B(r)$
  are even functions if $m = 0$, otherwise $B(r)$ loses this symmetry.
  Asymptotic flatness at $r = +\infty$ implies
\beq
  	2bc = -3\pi m,                                     \label{as-flat}
\eeq
  where $m$ is the Schwarzschild mass defined in the usual way.

  Under this asymptotic flatness assumption, for $m = c = 0$, we obtain the
  simplest symmetric configuration, the Ellis \wh\ \cite{h_ellis}: $A \equiv
  1$, $V \equiv 0$. With $m < 0$, we obtain a \wh\ with an AdS metric at the
  far end, corresponding to the cosmological constant $V(\phi)\big|_{r\to
  -\infty} = V_- < 0$. Further on, such configurations will be referred to
  as M-AdS \whs\ (where M stands for Minkowski, the flat asymptotic).
  Assuming $m > 0$, at large negative $r$ we obtain negative $A(r)$, such
  that $|A(r)| \sim R^2(r)$, and a potential tending to $V_- > 0$. Thus it
  is a regular \bh\ with a de Sitter asymptotic behavior far beyond the
  horizon, precisely corresponding to the above description of a black
  universe.

  In black-universe solutions, the horizon radius depends on both parameters
  $m$ and $b$ ($\min R(r)=b$), which also plays the role of a scalar charge
  since $\phi/\sqrt{2} \approx \pi/2 -b/r$ at large $r$. Since $A(0) = 1 +
  c$, the minimum of $R(r)$, located at $r = 0$, occurs in the R region if
  $c > -1$, i.e., if $3\pi m < 2b$ (it is then a throat, like that in \whs),
  right at the horizon if $c=-1$ (i.e., $3\pi m = 2b$) and in the T region
  beyond it if $ c < -1$, that is, $3\pi m > 2b$. It is then not a throat,
  since $r$ is a time coordinate, but a bounce in one of the scale factors
  $R(r)$ of the \KS\ cosmology; the other scale factor is $A(r)$.

  Let us mention that another important case of the system
  (\ref{Lagrangian}), the one with $V \equiv 0$, has already been studied in
  a number of papers. In this case we are dealing with Fisher's famous
  solution (\cite{Fisher}, 1948) for a canonical massless scalar in \GR\
  ($\ep = +1$) and three branches of its counterpart for $\ep = -1$,
  sometimes called the anti-Fisher solution, found for the first time by
  Bergmann and Leipnik \cite{BerLei} in 1957 and repeatedly rediscovered
  afterwards (as well as Fisher's solution). In the latter case the solution
  consists of three branches, one representing \whs\ \cite{kb73, h_ellis},
  the other two also containing throats but with singularities or horizons
  of infinite area at the far end instead of another spatial infinity
  \cite{kb73, cold-08}. An instability of Fisher's solution under \sph\
  perturbations was established long ago \cite{br-hod}, a similar
  instability of the \wh\ branch was discovered in  \cite{Gonzalez:2008wd,
  Gonzalez:2008xk} and the same for the other two branches in
  \cite{Bronnikov:2011if}. The case of zero mass in the anti-Fisher solution
  represents the Ellis \wh\ and is common with the solution under study,
  (\ref{sol-pha1}), $m=0$.

  In what follows, we will assume $m \ne 0$ and use the mass $m$ (which has
  the dimension of length and is equal to half the Schwarzschild radius in
  the units employed) as a natural length scale, putting, for convenience,
  $|3m| = 1$. Then the constant $c$ is used as a family parameter while $b$
  is found from the relation (\ref{as-flat}).

  We should remark that in all particular examples to be tested here for
  stability the background electromagnetic field is zero, but the general
  perturbation formalism is developed in Sec III for axial perturbations
  of systems with non-zero $F_{\mu\nu}$ as well. Actually in Sec IV non-zero
  $F_{\mu\nu}$ is also allowed, simply because the monopole perturbations do
  not excite an electromagnetic field in a spherically symmetric background.
  Perturbations of the electromagnetic field appear in higher multipoles of
  polar modes. The general formalism developed for non-zero $F_{\mu\nu}$ is
  intended to be used in further studies of systems with both scalar and
  electromagnetic fields.

\section{Linear axial perturbations: general analysis}

  In our analysis of axial perturbations, we use Chandrasekhar's notations:
  $x^0 =t$, $x^1 = \phi$, $x^2 = r$, $x^3 =\theta$, so that the coordinates
  along which the background has Killing vectors are enumerated first.
  Following Chandrasekhar \cite{Chandra}, we consider the metric
  (\ref{metric}) as a special case of the metric
\bearr                                                     \label{ds-Cha}
  	d s^2  = e^{2 \nu} d t^2  - e^{2 \psi} (d \phi
	- \sigma d t - q_{2} d r - q_{3} d \theta)^2
\nnn \inch
	- e^{2 \mu_{2}} d r^2  - e^{2\mu_{3}} d \theta^2 .
\ear
  Thus in (\ref{metric})
\bearr
	e^{2 \nu} = A(r), \qquad\ \  e^{2\mu_{2}} = A^{-1}(r),
\nnn
	e^{2\mu_{3}} = R(r)^2 , \qquad e^{2\psi} = R(r)^2 \sin^2 \theta.
\ear

  The background electromagnetic field is taken in the form
\beq
	F_{02} = -Q_{*}/R(r)^2 ,
\eeq
  that is, only a radial electric field, and $Q_*$ is the (effective) charge.

  All calculations and results can be easily rewritten for magnetic fields,
  with nonzero $F_{13}$, owing to the Maxwell field duality. One can bear
  in mind that configurations like \whs\ and black universes can possess
  electric or magnetic fields without any real electric charges or magnetic
  monopoles, due to their geometry, actually realizing Wheeler's concept of
  a ``charge without charge''.

  It is easy to see that axial perturbations of a scalar field vanish.
  Then, $\sigma$, $q_2$ and $q_3$ are perturbed, while $\psi$,
  $\mu_2$, $\mu_3$ and $\nu$ remain unperturbed.

  The axial gravitational perturbations obey the equations
\beq
	\delta R_{13} =0, \qquad \delta R_{12} = 2 Q_{*} R^{-2} F_{01}
\eeq

  Let us introduce new variables:
\beq\label{-1}
	Q_{ik} = q_{i, k} -q_{k, i}, \qquad Q_{i0} = q_{i, 0} -\sigma_{,i},
\eeq
  with $i, k = 2, 3$, and
\beq
	E \equiv F_{01} \sin \theta.
\eeq
  Recall that we use the numbers $0, 1, 2, 3$ for $t$, $\phi$, $r$
  and $\theta$ coordinates, respectively. The Maxwell equations, subject to
  only first-order perturbations, have the form
\bearr
	(e^{\psi + \mu_2} F_{12})_{,3} + (e^{\psi + \mu_3} F_{31})_{,2} = 0,
\yyy
	(e^{\psi + \nu} F_{01})_{,2} + (e^{\psi + \mu_2} F_{12})_{,0} = 0,
\yyy
	(e^{\psi + \nu} F_{01})_{,3} + (e^{\psi + \mu_3} F_{13})_{,0} = 0,
\yyy
	(e^{\mu_2 + \mu_3} F_{01})_{,0} + (e^{\nu + \mu_3} F_{12})_{,2}
\nnn \cm
 	+ (e^{\nu + \mu_2} F_{13})_{,3} = e^{\psi + \mu_3} F_{02} Q_{02}.
\ear

  After some algebra the Maxwell equations can be written in the form
\bearr
	R e^{-\nu} E_{,0, 0} - (e^{2 \nu} (R e^{\nu} E)_{,r})_{,r}
	+ \frac{e^{\nu}}{R} \sin \theta \left(E_{,\theta}
				\frac{1}{\sin\theta}\right)_{, \theta}
\nnn \inch \label{0}
	= - Q_{*} (\sigma_{, 2, 0} - q_{2, 0, 0}) \sin^2  \theta
\ear

  From $\delta R_{13} =0$ and $\delta R_{12} = 2 Q_{*} R^{-2} F_{01}$
  it follows
\beq\label{1}
	(e^{3 \psi + \nu - \mu_2 - \mu_3} Q_{23}),_{2}
		- (e^{3 \psi - \nu + \mu_2 - \mu_3} Q_{03}),_{0} =0
\eeq
  and
\bearr
	(e^{3 \psi + \nu - \mu_2 - \mu_3} Q_{23}),_{3}
		- (e^{3 \psi - \nu - \mu_2 + \mu_3} Q_{02}),_{0}
\nnn \inch 							\label{2}
 		= e^{2 \psi + \nu + \mu_3} Q_{*} R^{-2} F_{01}.
\ear
  After introducing the new function
\beq
		Q \equiv R^2 A Q_{23} \sin^{3} \theta,
\eeq
  \eqs (\ref{1}) and (\ref{2}) can be written in the form
\bearr\label{3}
	\frac{A}{R^2 \sin^3 \theta} \frac{\d Q}{\d r}
			= \sigma_{, 3, 0} - q_{3, 0, 0},
\yyy\label{4}
	\frac{A}{R^4 \sin^3 \theta} \frac{\d Q}{\d \theta}
	 = -\sigma_{, 2, 0} + q_{2, 0, 0}
	 	+ \frac{4 Q_{*} e^{\nu} E}{R^2 \sin^2  \theta}.
\ear
  Let us differentiate \eq (\ref{3}) in $r$ and \eq (\ref{4}) in $\theta$
  and then add the results. After some algebra we have
\bearr
	R^4 \frac{\d}{\d r} \left(\frac{A}{R^2} \frac{\d Q}{\d r}\right)
	+ \sin^{3} \theta \frac{\d}{\d \theta} \left(\frac{1}{\sin^{3}
	\theta} \frac{\d Q}{\d \theta}\right) - \ddot{Q} R^2 e^{- 2 \nu}
\nnn  \cm							\label{5}
 	= 4 Q_{*} e^{\nu} R \sin^{3} \theta \frac{\d}{\d \theta}
 		\left(\frac{E}{\sin^2  \theta}\right).
\ear

  Now let us return to the Maxwell field perturbation equation (\ref{0}).
  Using (\ref{4}), we can get rid of the term containing $\sigma_{, 2, 0} -
  q_{2, 0, 0} $ in (\ref{-1}). After some algebra and using the relations
  $E\sim e^{i \omega t}$,  $Q \sim e^{i \omega t}$, we obtain
\bearr
	(e^{2 \nu} (R e^{\nu} E)_{, r})_{, r} + E (\omega^2 R e^{-\nu}
		- 4 Q_{*}^2  e^{\nu} R^{-3})
\nnn\ \ 							\label{6}
   - e^{\nu} R^{-1} \sin \theta \left(E_{,\theta} \frac{1}{\sin\theta}\right)_{, \theta}
	= - \frac{Q_{*}}{R^4 \sin \theta} \frac{\d Q}{\d \theta}.
\ear

  The angular variable can be separated by the following ansatz:
\bearr
	Q(r, \theta) = Q(r) C_{\ell + 2}^{-3/2}(\theta),
\yyy
	E(r, \theta) = \frac{E(r)}{\sin \theta} \frac{d C_{\ell + 2}^{-3/2}
		(\theta)}{d \theta} = 3 E(r) C_{\ell + 1}^{-1/2}(\theta),
\ear
  where $C_a^b$ are Gegenbauer polynomials.
  \eqs (\ref{5}) and (\ref{6}) then read
\bearr								\label{7}
	\Delta \frac{d}{d r}\!\Big(\frac{\Delta}{R^4} \frac{d Q}{dr} \Big)
	- \mu^2 \frac{\Delta}{R^4} Q + \omega^2 Q = - \frac{4 Q_{*} \mu^2
		\Delta e^{\nu} E}{R^3},
\yyy
	(e^{2 \nu} (R e^{\nu} E)_{,r})_{,r} - (\mu^2 + 2) e^{\nu} R^{-1} E
\nnn \cm 							\label{8}
    +E (\omega^2 R e^{-\nu} - 4 Q_{*}^2  e^{\nu} R^{-3}) = - Q_{*}Q R^{-4},
\ear
  where $\Delta = R^2 e^{2 \nu}$, $\mu^2 = (\ell-1)(\ell + 2)$, and we use
  the tortoise coordinate $r_*$ defined by
  $d/d r_* = \Delta R^{-2} d/d r$.

  After passing over from $Q$ and $E$ to the new functions $H_1$ and
  $H_2$ using the relations
\beq
	Q= R H_2, \qquad Re^{\nu} E = - \frac{H_1}{2 \mu},
\eeq
  \eqs (\ref{7}) and (\ref{8}) can be reduced to
\bearr								\label{9}
 	\Lambda^2 H_2 = \left(-\frac{R_{, r_* r_*}}{R} +
	\frac{2 R_{, r_*}^2}{R^2} + \mu^2 \frac{\Delta}{R^4}\right) H_2
		 + \frac{2 Q_* \mu \Delta}{R^5} H_1,
\nnn
\yyy 								\label{10}
	\Lambda^2 H_{1} = \frac{\Delta}{R^4} \left((\mu^2 + 2)
		+ \frac{4 Q_*^2}{R^2 }\right) H_1 
		 + \frac{2 Q_* \mu \Delta}{R^5} H_2,
\ear
  where we have introduced the operator
\beq
	\Lambda^2 = \frac{d^2}{d r_{*}^2} + \omega^2.
\eeq
  For an electrically neutral background $Q_{*}=0$, $H_1 =0$, and
  \eq (\ref{9}) reduces to the Schr\"odinger-like equation
\beq							\label{wavelike}
	\frac{d^2  H_2}{d r_{*}^2} + (\omega^2  - \Veff(r))H_2 =0
\eeq
  with the effective potential
\beq							\label{potential}
	\Veff(r) = A(r) \frac{(\ell {+} 2)(\ell {-}1)}{R^2} +
			R(R^{-1})_{,r_{*}, r_{*}},
\eeq

  As a partial verification of the above relations, we can check that \eqs
  (\ref{9}) and (\ref{10}) reproduce some known special cases. Thus, \eq
  (\ref{potential}), obtained here in the Chandrasekhar approach
  \cite{Chandra}, coincides with \eq (4.13) of \cite{Lechtenfeld}, obtained
  in the Regge-Wheeler approach. In addition, \eqs (\ref{9}), (\ref{10})
  coincide with \eqs (144), (145), p. 230 of \cite{Chandra} in the \RN\
  limit, and, as a result, the potential (\ref{potential}) coincides with the
  well-known Regge-Wheeler potential for Schwarzschild black holes.

\section{Polar perturbations}

  For polar perturbations let us consider the Einstein-scalar equations
  following from (\ref{Lagrangian}) with $F\mn =0$
\bearr                                            	\label{E-sca}
	R_{\mu \nu} + \ep \d_{\mu}\Phi \d_{\nu} \Phi
			- \delta_{\mu \nu} V(\phi) =0,
\nnn
   \ep \Box \Phi + V_{\phi} = 0, \qquad V_{\phi} \equiv dV/d\phi,
\ear
  where $\Box = \nabla^\alpha \nabla_\alpha$ is the d'Alembert operator. In
  the polar case we can put $\sigma = q_{2} = q_{3} = 0$, while $\delta
  \nu$, $\delta \mu_{2}$, $\delta \mu_{3}$, and $\delta \psi$ do not vanish.
  In addition, we perturb the scalar field,
\[
	\phi(r, t) = \phi(r) + \delta \phi(r, t).
\]
  Let us restrict ourselves to the lowest frequency modes, corresponding to
  spherically symmetric (or radial) perturbations, $\ell =0$. This gives
  $\delta \psi = \delta \mu_{3}$. Then we can use the gauge freedom and
  fix\footnote
	{See more details on gauges and gauge-invariant perturbations in
	\cite{Gonzalez:2008wd, Bronnikov:2011if}. The present notations
	are related to those in \cite{Bronnikov:2011if} as follows:
   \[
      r \mapsto u, \ \ \nu \mapsto \gamma, \ \ R(r) \mapsto r(u),\ \
      	\mu_2 \mapsto \alpha, \ \ \mu_3 \mapsto \beta.
   \]
	}
\beq
	\delta \psi = \delta \mu_3 \equiv 0 \ \then \ \delta R(r,t) =0.
\eeq

  Tedious calculations allow us to reduce the perturbation equations to a
  single wave equation
\beq                                                         \label{wave}
	e^{-2 \mu_2 - 2 \nu} \delta \ddot{\phi} - \delta \phi'' - \delta\phi'
		(\nu'  - 2 \mu_{3}'  + \mu_{2}' ) + U \delta \phi =0,
\eeq
  where
\beq                                                         \label{U}
	U \equiv e^{- 2 \mu_2}\left(\ep(e^{2 \mu_3}-V) \frac{(\phi' )^2}
		{(\mu_{3}' )^2} - \frac{2 \phi' }{\mu_{3}'} V_{,\phi}
		+ \ep V_{,\phi \phi}\right).
\eeq
  Introducing the new function $\Psi$ by putting
\beq
	\delta \phi = \Psi e^{\mu_3 + i \omega t},
\eeq
  we bring the wave equation to the Schr\"odinger-like form
\beq                                                         \label{Schr}
	\frac{d^2 \Psi}{d r_{*}^2 } + (\omega^2
				- \Veff(r_{*}))\Psi = 0,
\eeq
  with the effective potential
\bearr                                                      \label{Veff1}
	\Veff = U + \frac{1}{R}\frac{d^2 R}{dr_{*}^2 }.
\yyy 							\label{UB}
	\frac{U}{A}  = \frac{\ep \phi'^2}{R'^2}(R^2 V - 1)
 	+ \frac{2 \phi' R V_{,\phi}}{R'} + \ep V_{,\phi ,\phi}.
\ear

\section{Stability analysis}

\subsection{Methods}

\noi
{\bf Finite potentials.} If the effective potential $\Veff$ is finite and
  positive-definite, the differential operator
\beq
	W = - \frac{d^2 }{d r_{*}^2} + \Veff
\eeq
  is a positive self-adjoint operator in $L^2 (r_{*}, d r_{*})$,
  the space of functions satisfying proper boundary conditions.
  Then $W$ has no negative eigenvalues, in other words,
  there are no normalizable solutions to the corresponding \Schr\ equation
  with well-behaved initial data (smooth data on a compact support) that
  grow with time, and the system under study is then stable under this
  particular form of perturbations. Therefore, in our stability studies,
  our main concern will be regions where the effective potentials are
  negative since possible instabilities are indicated by such regions.

  The response of a stable black hole or wormhole to external perturbations
  is dominated at late times by a set of damped oscillations, called
  \emph{quasinormal modes} (QNMs). Quasinormal frequencies do not depend on
  the way of their excitation but are completely determined by the
  parameters of the configuration itself. Thus quasinormal modes form a
  characteristic spectrum of proper oscillations of a black hole or a
  wormhole and could be called their ``fingerprints''. Apart from black hole
  physics, QNMs are studied in such areas as gauge/gravity correspondence,
  gravitational wave astronomy \cite{Konoplya:2011qq,Kokkotas:1999bd} and
  cosmology \cite{QNMcosmology}.

  The quasinormal boundary conditions for black hole perturbations imply pure
  incoming waves at the horizon and pure outgoing waves at spatial infinity.
  For asymptotically flat solutions, the quasinormal boundary conditions are
\beq                                                       \label{Psi-flat}
	\Psi \propto e^{ \pm i \omega r_{*}}, \quad r_{*} \to \pm \infty.
\eeq

  The proper oscillation frequencies correspond in a sense to a ``momentary
  perturbation'', that is, to the situation where one looks at a response of
  the system (say, a wormhole) to initial perturbation when the source of
  the perturbation stopped to act. This is the essence of the word
  ``proper''. Thus, in the wormhole case, no incoming waves are allowed
  coming from either of the infinities. The issue of the boundary conditions
  for quasinormal modes of wormholes is not completely new and was
  considered in \cite{Konoplya-Molina}.

  Therefore, for wormholes the condition ``pure incoming waves at the
  horizon'' is replaced with ``pure outgoing waves at the other spatial
  infinity'' \cite{Konoplya-Molina, Konoplya:2010kv}. For asymptotically
  anti-de Sitter black holes or \whs, the AdS boundary creates an effective
  confining box \cite{Konoplya:2011qq}, so that on the AdS boundary one
  usually requires the Dirichlet boundary conditions
\beq
	\Psi \to 0, \quad r_{*} \to \infty.
\eeq
  This choice is not only dictated by the asymptotic of the wave equation at
  infinity but is also consistent with the limit of purely AdS space-time:
  the QNMs of an AdS black hole approach the normal modes of empty AdS
  space-time in the limit of a vanishing black hole radius
  \cite{Konoplya:2002zu}. Thus quasinormal modes of a compact object (a
  black hole or wormhole) in AdS space-time look like normal modes of empty
  AdS space-time ``perturbed'' by the compact object.

  If the effective potential is negative in some region, growing quasinormal
  modes can appear in the spectrum, indicating an instability of the system
  under such perturbations. It turns out that some potentials with a
  negative gap still do not imply instability. If there are no growing
  quasinormal modes in the black-hole or wormhole spectrum, this object is
  stable against linear perturbations.

\medskip\noi
{\bf Singular potentials and their regularization.}
  It can be seen from (\ref{Veff1}) (provided that $VR^2 < 1$) that the
  effective potential $\Veff$ for radial perturbations forms an infinite
  wall located at a throat, with the generic behavior $\Veff \sim
  (r-r_{\rm throat})^{-2}$, since we have there $R'(r) \sim r-r_{\rm
  throat}$.  As a result, perturbations are independent at different sides
  of the throat, necessarily turn to zero at the throat itself, and we thus
  lose part of the information on their possible properties. To remove the
  divergence, one can use a method on the basis of S-deformations as
  described in \cite{Volkov:1998cc,Kodama:2003kk} for solutions with
  arbitrary potentials $V(\phi)$. For anti-Fisher \whs\ ($V(\phi)\equiv 0$)
  it was applied in \cite{Gonzalez:2008wd}.

  If the scalar potential $V(\phi)$ is zero, the effective potential
  (\ref{Veff1}) takes the form
\beq
      \Veff(r) = \frac{A} {R^2}-\frac{A^2R'{}^2}{R^2}
  	-\frac{\ep A\phi'{}^2}{R'{}^2}.
\eeq
  In this case the general static solution ($\omega=0$) to \eq (\ref{Schr})
  is
\bearr	\nqq						\label{statsol}
     \Psi_0 (r) = C_1 R \left(1-\frac{\ep A R \phi'{}^2}{A' R'}\right)
\nnn \ \
	+ \frac{C_2}{R}\left[\frac{\phi}{A'}-\frac{\ep A R\phi'{}^3}
	  {A'{}^2 R'}\left(\frac{\phi}{\phi'}-\frac{2A}{A'}\right)\right],
\ear
  and its special cases with $C_1=0$ and $C_2=0$ were used in
  \cite{Bronnikov:2011if, Gonzalez:2008wd} to remove the divergence in
  $\Veff$. Indeed, we can introduce the new wave function
\beq                                                     \label{o-Psi}
     \overline{\Psi}=S\Psi-\frac{d\Psi}{dr_*},
     \quad \mbox{where} \quad S=\frac{1}{\Psi_0}\frac{d\Psi_0}{dr_*},
\eeq
  which satisfies the equation
\beq                                                     \label{o-Schr}
	\frac{d^2\overline{\Psi}}{d r_{*}^2 } + \left(\omega^2
	 - \overline{V}_{\rm eff}(r_*)\right)\overline{\Psi} =0
\eeq
  with the effective potential
\beq                                                     \label{o-Veff}
	\overline{V}_{\rm eff}
	= 2S^2 - \Veff,
\eeq
  and the new effective potential is everywhere regular if either $C_1=0$ or
  $C_2=0$. This transformation was used to prove the instability of
  anti-Fisher \whs\ \cite{Gonzalez:2008wd} and other anti-Fisher solutions
  \cite{Bronnikov:2011if},

\begin{figure*}      
\resizebox{0.5\linewidth}{!}{\includegraphics*{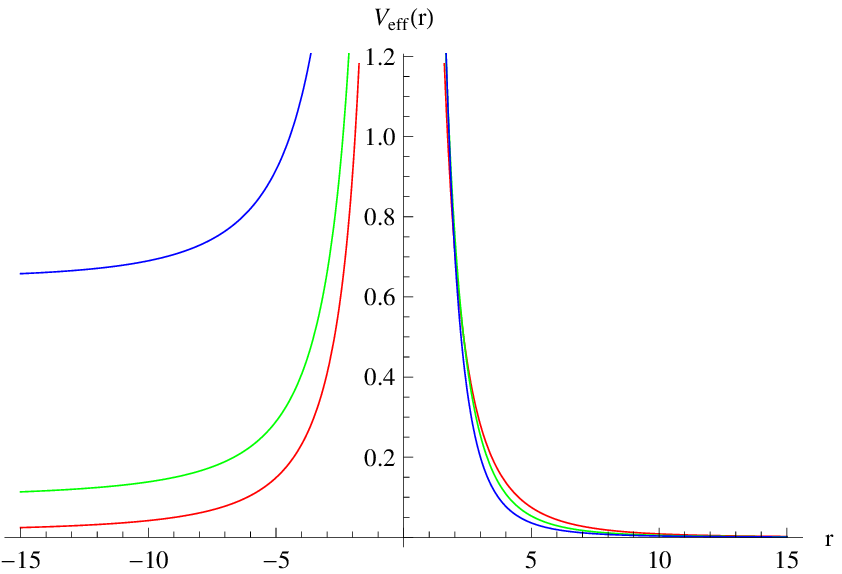}}%
\resizebox{0.5\linewidth}{!}{\includegraphics*{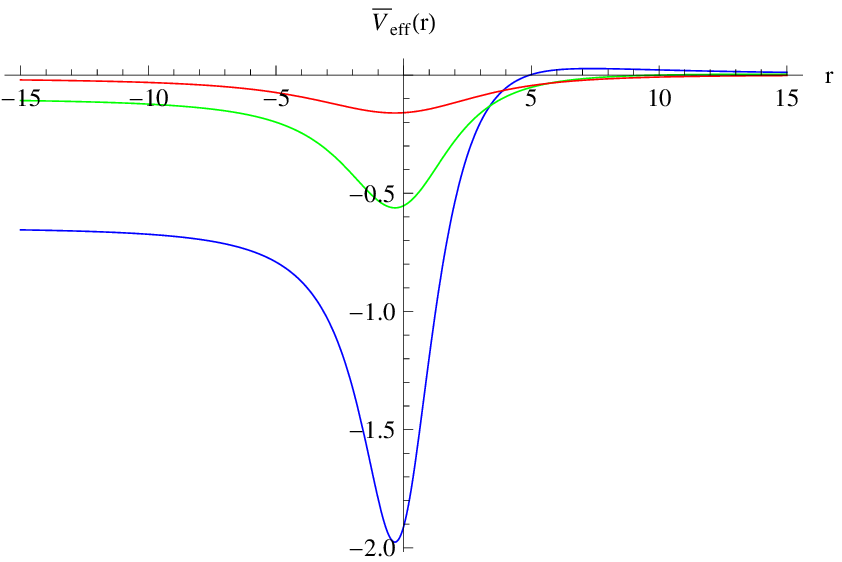}}
\caption{Effective potentials for spherically symmetric perturbations of
      M-AdS wormholes with $m=-1/3$, $c=0.3$ (red), $c=0.5$ (green), $c=0.8$
      (blue) with divergences at the throat (left figure) and the
      corresponding regular ones found numerically (right figure).  Larger
      values of $c$ correspond to larger absolute values of $\Veff$ at the
      AdS boundary (they are positive for divergent potentials and negative
      for regular ones).}\label{_fig1}
\end{figure*}

\begin{figure}       
\resizebox{\linewidth}{!}{\includegraphics*{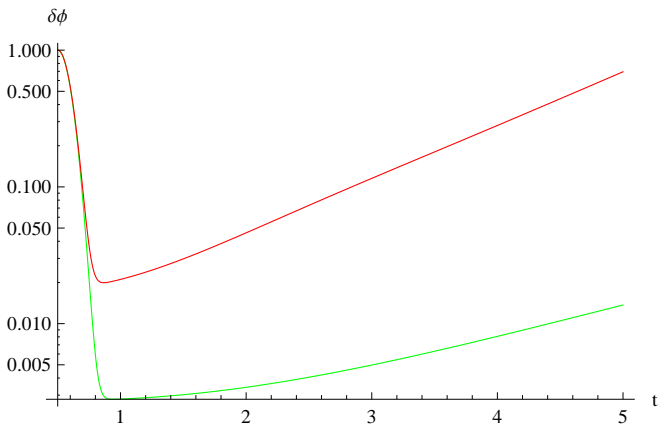}}%
\caption{Time-domain evolution of spherically symmetric perturbations of
	M-AdS wormholes with $m=-1/3$, $c=0.5$ (green, bottom), and $c=0.8$
	(red, top).}\label{_fig2}
\end{figure}

  It turns out that $C_1$ and $C_2$ can be both nonzero, leading to
  regularized effective potentials with the same quasinormal spectrum,
  though not preserving the symmetry $\phi\leftrightarrow -\phi$. This
  means that we can fix the boundary conditions arbitrarily and, if we find
  a ``good'' static solution, we can use it to remove the singularity of the
  effective potential.

  For nonzero scalar potentials $V(\phi)$, analytical expressions for static
  perturbations are unknown. Therefore, such suitable solutions must be found
  numerically under proper boundary conditions.

  First we notice that since $\Psi_0$ is a static solution to \eq
  (\ref{Schr}), $S$ satisfies the Riccati equation
\beq					\label{Seq}
	\frac{dS}{dr_*}+S^2-V_{\rm eff}=0.
\eeq
  Substituting (\ref{Veff1}) into (\ref{Seq}), we find an expansion for $S$
  near the throat as \cite{Bronnikov:2011if}
\bearr\label{Sexp}
	S(r) = -\frac{1+c}{r} - \frac{4c^2}{3m\pi^2}
                	+ \frac{4c^2(4c^2+(1+c)\pi^2)}{9m^2(1+c)\pi^4}r
\nnn \inch
	+K r^2 + \ldots,
\ear
  where $K$ is an arbitrary constant.

  With this expression we find the boundary condition for the function
  $S(r)$ at some points close to the throat on both sides. Then we integrate
  (\ref{Seq}) numerically and find $S(r)$ between the throat and both
  asymptotical regions. Having $S(r)$ at hand, we find the regularized
  effective potential (\ref{o-Veff}); it is finite at the throat due to
  (\ref{Sexp}).

  We assigned different values to the free constant $K$. As a rule, we were
  able to integrate \eq (\ref{Seq}) numerically in a sufficiently wide
  range of $r$ near the throat. The regularized effective potentials
  obtained in this way always lead to the same growth rate of the
  perturbation function. However, for some particular values of $K$, the
  numerical integration with the Wolfram Mathematica\textregistered{}
  built-in function encounters a growing numerical error. Apparently,
  in these cases some alternative methods of integration should be used.

  In addition, we used the same method to find numerically the regularized
  potential for the Branch B anti-Fisher solution whose stability properties
  had been studied previously \cite{Bronnikov:2011if}. Although the
  numerically found potential differs from \eq (68) of
  \cite{Bronnikov:2011if}, the time-domain profile again shows the same
  growth rate. This confirms the correctness of the method suggested and
  used here.

\subsection{M-AdS wormholes with negative mass}

\medskip\noi
{\bf Polar perturbations.}
  Now we are in a position to apply the above method to the special case of
  M-AdS \whs\ (\ref{sol-pha1}) with $m < 0$, which are asymptotically AdS
  as $r\to - \infty$. In Fig.\,\ref{_fig1} one can see that although
  the initial (singular) effective potentials $\Veff$ are positive definite,
  the regularized potentials $\overline{V}_{\rm eff}$ are negative in some
  range around the throat and at the AdS boundary. This behavior usually
  indicates an instability but does not guarantee it \cite{meAdSstability}.
  Therefore, to prove the instability of M-AdS wormholes, we have used the
  time-domain integration method proposed by Gundlach et al.
  \cite{Gundlach:1993tp} and later used by a number of authors (see, e.g.,
  \cite{timedomain}). The method shows a convergence of the time-domain
  profile with diminishing the integration grid and increasing the precision
  of all computations. We imposed the Dirichlet boundary conditions at the
  AdS boundary, as described in \cite{Wang:2000dt}. Fig.\,\ref{_fig2} shows
  examples of time-domain profiles for the evolution of perturbations.  The
  growth of the signal allows us to conclude that such wormholes are
  unstable against radial perturbations. At larger $c$ we found the
  regularized potential with deeper negative wells and observed a quicker
  growth of the signal. Therefore we conclude that all such M-AdS wormholes
  are unstable.

\begin{figure*}            
\resizebox{0.5\linewidth}{!}{\includegraphics*{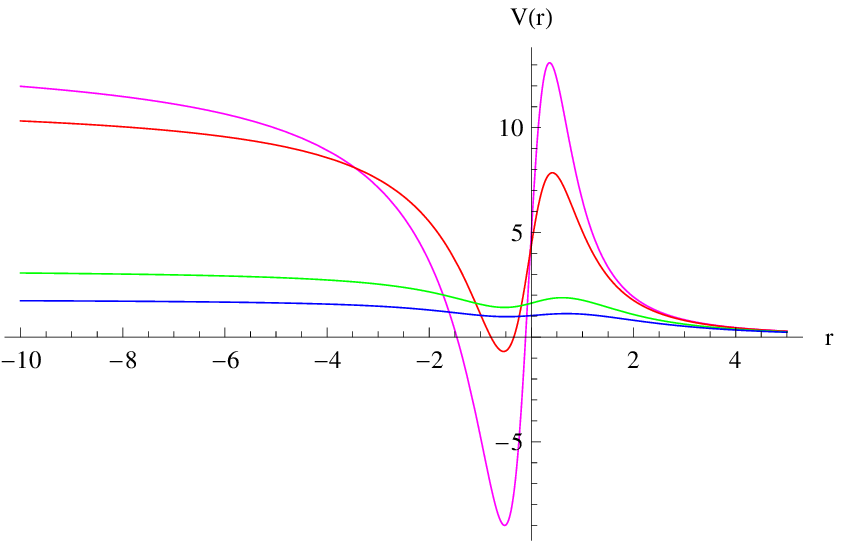}}%
\resizebox{0.5\linewidth}{!}{\includegraphics*{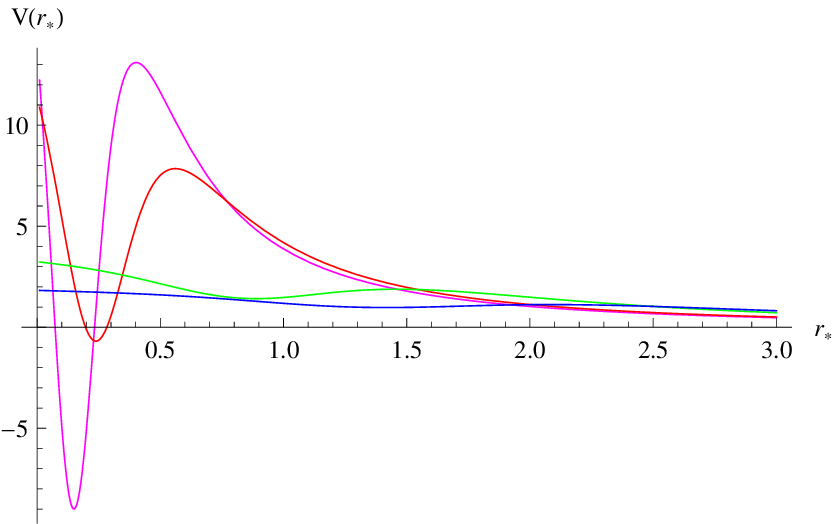}}
\caption{Effective potential as a function of the radial coordinate $r$
  (left figure) and of the tortoise coordinate $r_*$ (right figure) for axial
  gravitational perturbations of M-AdS wormholes with $m=-1/3$ for $\ell =2$,
  $c=0.8$ (no negative gap), $1.0$, $1.8$ (a negative gap), $2.2$
  (a deep negative gap).
	}   \label{_fig3}
\end{figure*}

\begin{figure*}            
\resizebox{.5\linewidth}{!}{\includegraphics*{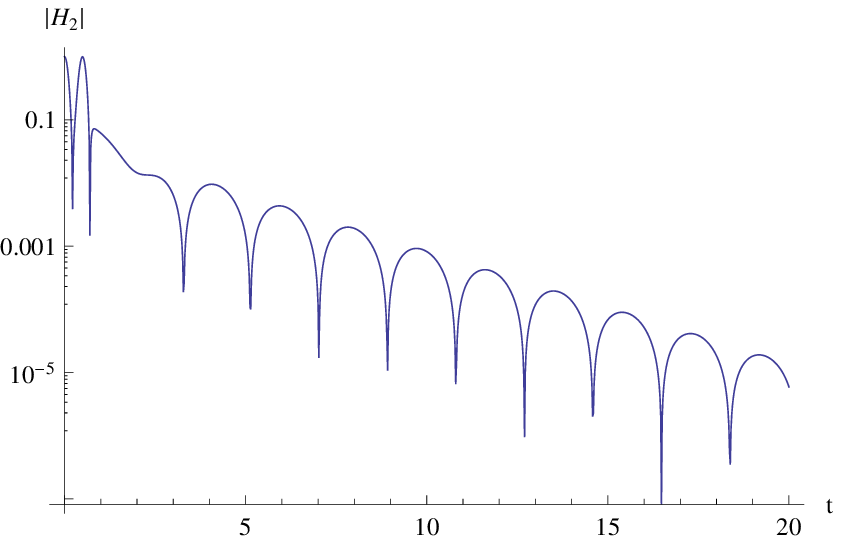}}%
\resizebox{.5\linewidth}{!}{\includegraphics*{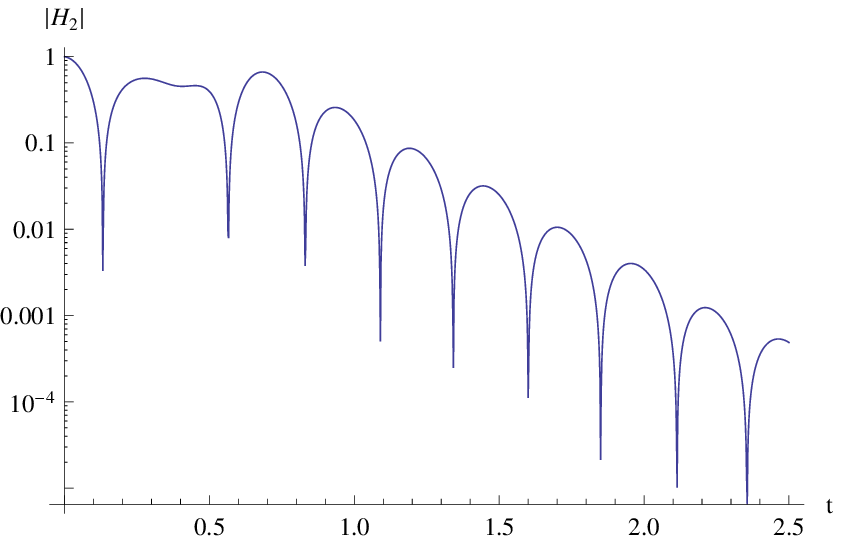}}
\caption{Time-domain profiles of the evolution of axial ($\ell =2$)
 	 perturbations of M-AdS wormholes ($m=-1/3$)
	 with $c=1$ (left) and $c=5$ (right).}
\label{fig_profiles}
\end{figure*}

\begin{figure*}          
\resizebox{\linewidth}{!}{\includegraphics*{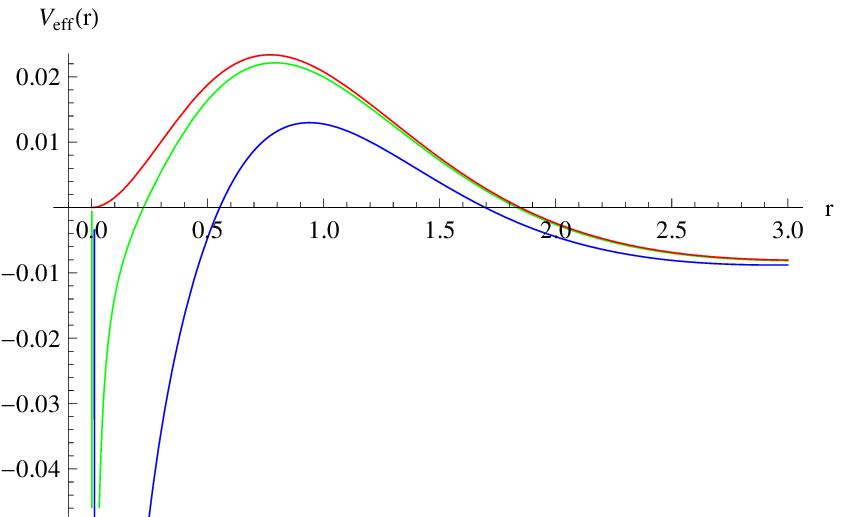}
\includegraphics*{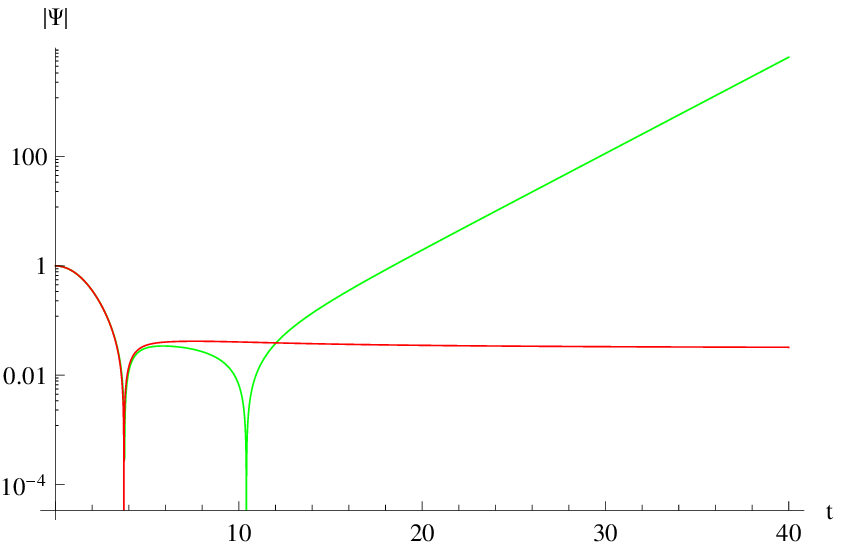}}
 \caption{Left panel: Effective potentials for radial perturbations of a
   black universe with $\ell = 0$, $m= 1/3$, $c=-1$ (red, top), $c=-1.001$
   (green), $c=-1.01$ (blue, bottom). The potentials vanish at the horizon.
   Right panel: time-domain perturbation evolution for $c=-1$ (red, stable),
   and $c=-1.001$ (green, unstable).}\label{potc=-1_fig}
\end{figure*}

\begin{figure}           
\resizebox{\linewidth}{!}{\includegraphics*{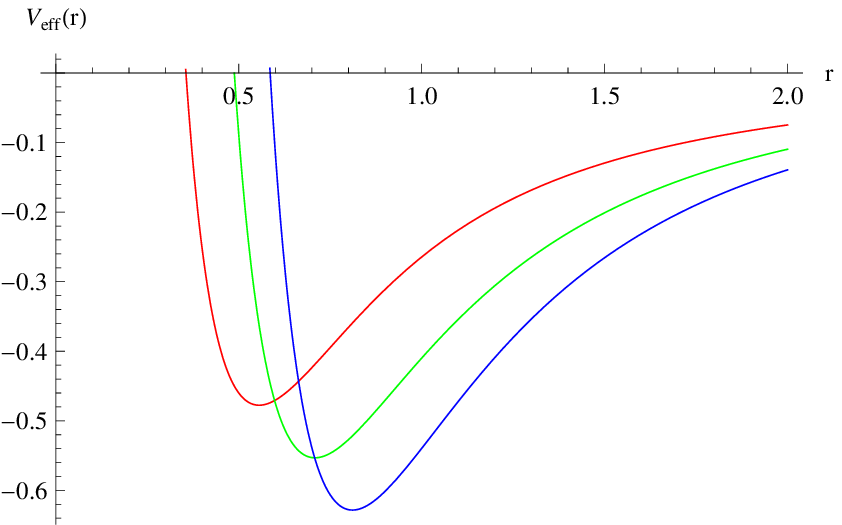}}
\caption{Effective potential for radial perturbations of a black universe
    with $\ell = 0$, $m= 1/3$, $c=-1.5$ (red, top), $c=-2$ (green), $c=-3$
    (blue, bottom). The potentials vanish at the horizon. }\label{potcn_fig}
\end{figure}

\medskip\noi
{\bf Axial perturbations.} The effective potentials $\Veff$ for axial
  perturbations of M-AdS wormholes $V_{\rm eff}$ are plotted in
  Fig.\,\ref{_fig3}. One can see that above some threshold value of $c$,
  $\Veff(r)$ has a negative gap. This threshold value of $c$ is
  $c \approx 1.737$ in units for which $m =-1/3$. A further increase of $c$
  makes the negative gap deeper, however, time-domain profiles for the
  evolution of axial perturbations show a decay of the signal without any
  indication of instability (see Fig.\,\ref{fig_profiles}).

\subsection{Black universes}

{\bf Polar perturbations.} Black universes correspond to the metric
  (\ref{metric}), (\ref{sol-pha1}) with $m > 0$, $c < 0$. Black universes
  with $c \leq -1$, or equivalently $3\pi m \geq 2b$, have no throat in the
  static region. From Fig.\,\ref{potc=-1_fig} one can see that for $c=-1$
  the effective potential has a negative gap, however, time-domain
  integration proves that in this case the black universe is stable. For
  smaller values of $c$, an additional negative gap appears between the peak
  and the horizon, leading to an instability even for $c=-1.001$. For
  large negative $c$ the potential peak disappears, and the potential
  becomes negative everywhere (see Fig.\,\ref{potcn_fig}), which inevitably
  creates an instability.

\begin{figure*}                  
\resizebox{0.5\linewidth}{!}{\includegraphics*{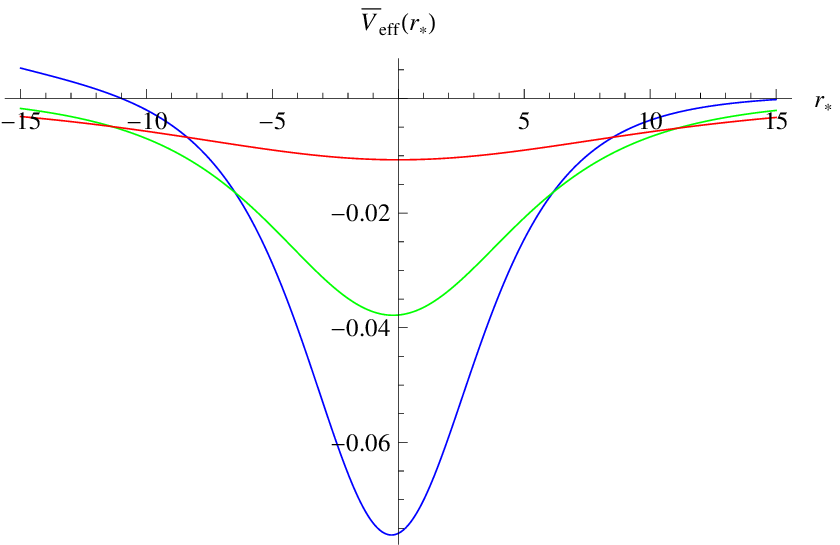}}%
\resizebox{0.5\linewidth}{!}{\includegraphics*{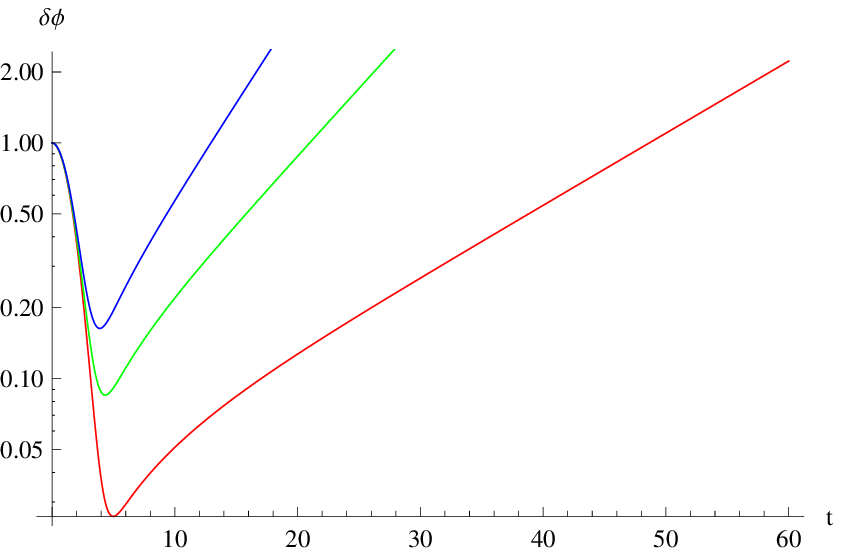}}
\caption{Regularized effective potentials (left panel, top to bottom) and
	time-domain profiles (right panel, bottom to top) for radial
	perturbations of a black universe with $\ell = 0$, $m= 1/3$,
	$c=-0.1$ (red), $-0.2$ (green), $-0.3$ (blue).}\label{BUtrans}
\end{figure*}

\begin{figure*}                  
\resizebox{0.5\linewidth}{!}{\includegraphics*{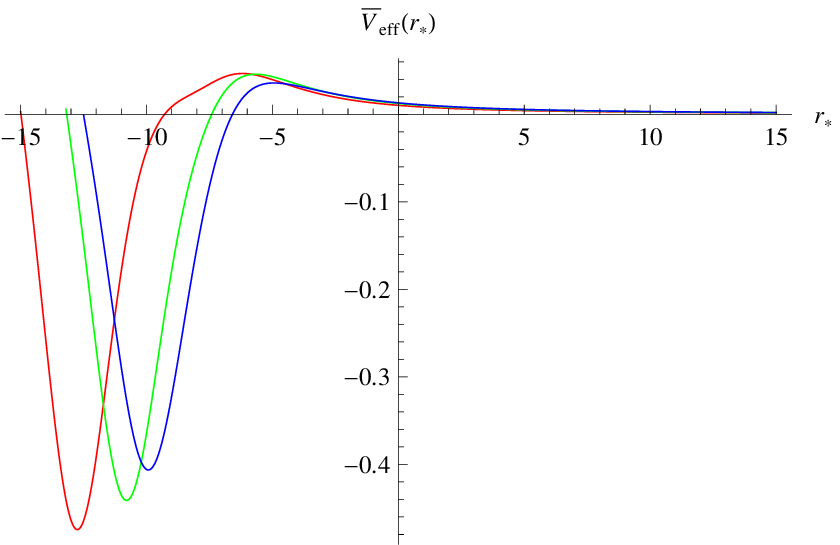}}%
\resizebox{0.5\linewidth}{!}{\includegraphics*{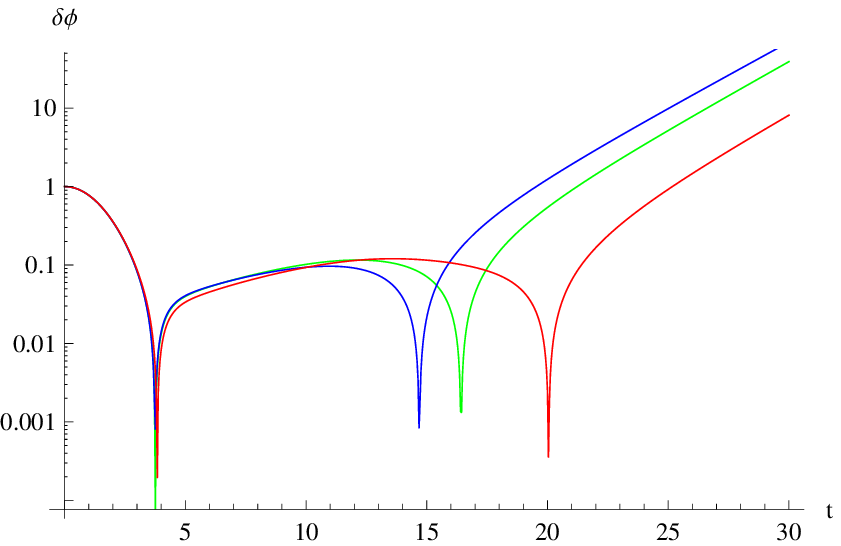}}
\caption{Regularized effective potentials (left panel) and time-domain
      profiles (right panel) for radial perturbations of a black universe
      with $\ell = 0$, $m= 1/3$, $c=-0.90$ (blue), $-0.95$ (green), $-0.99$
      (red). Larger negative values of $c$ correspond to deeper negative
      gaps and later growing phase of the signal.}\label{BUtranslim}
\end{figure*}

\begin{figure*}           fig 9
\resizebox{\linewidth}{!}{\includegraphics*{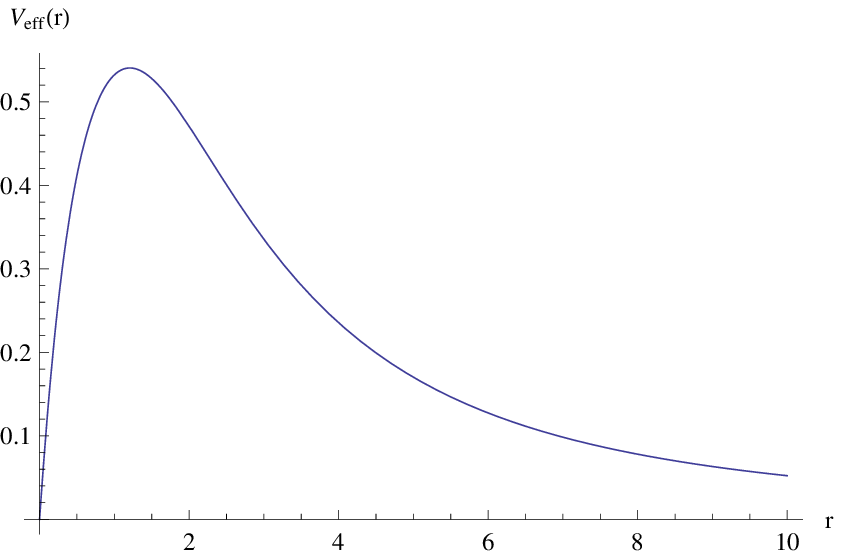}
	    \includegraphics*{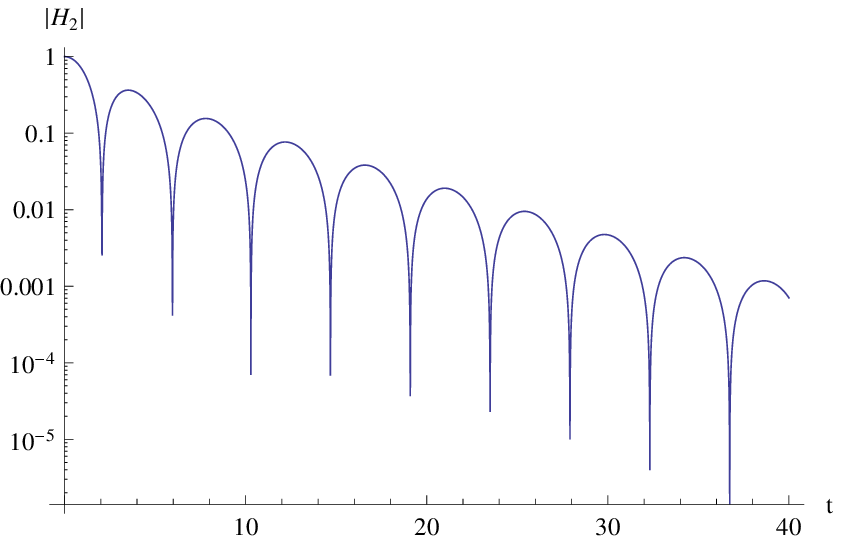}}
\caption{The effective potential (left figure) and the time-domain profile
	(right figure) for axial gravitational perturbations of black
	universes for $\ell =2$, $m=1/3$, $c=-1$. The potential vanishes at
	the horizon $r=0$.} \label{_fig4}
\end{figure*}

  At $c > -1$ the throat $r=0$ is in the static region. In this case we
  numerically find the regularized effective potentials which have large
  negative gaps leading to instabilities (see Figs.\,\ref{BUtrans},
  \ref{BUtranslim}). As $c$ approaches zero, the growth rate of time-domain
  profile decreases, still remaining positive because the effective potential
  remains negative in a wide region near the throat (Fig.\,\ref{BUtrans}).

  As $|c|$ grows the negative gap becomes deeper and narrower, giving way
  to a small positive hill, which becomes broader (see
  Fig.\,\ref{BUtranslim}) as $c$ approaches $-1$. However, we do not observe
  a decrease in the growth rate as expected when approaching the
  parametric region of stability near $c=-1$. This can be an indication that
  the only parameter for which a black universe can be stable is $c=-1$.

\medskip\noi
{\bf Axial perturbations and quasinormal oscillation frequencies.}  From
  Fig.\,\ref{_fig4} we can see that in the static region the effective
  potential is positive-definite. Therefore, if we perturb a black hole
  ``on the right'' of the event horizon (Fig.\,\ref{_fig4}), such
  perturbations are stable. Beyond the event horizon, in the cosmological
  region, the effective potential can take large negative values, but this
  has no effect on perturbations propagating outside the horizon. We
  therefore conclude that the black universes under consideration are stable
  against axial perturbations in the static region.

  As we have shown, black universes at $c=-1$ ($b=3\pi m/2$) are stable
  against polar monopole perturbations, and their response to external
  perturbations is dominated at late times by the quasinormal
  (QN) frequencies.  Supposing $\Psi \propto e^{-\imo \omega t}$,
  quasinormal modes can be written in the form
\[
	\omega = \omega_{\re} - \imo\omega_{\im},
\]
  where a positive $\omega_{\im}$ is proportional to the decay rate of a
  damped QN mode. The low-lying axial QN frequencies have the smallest
  decay rates in the spectrum and thus dominate in a signal at sufficiently
  late times. They can be calculated with the help of the WKB approach
  \cite{WKB, WKBorder}.

  Introducing $Q = \omega^2 - V_{\rm eff}$, the 6-th order WKB formula reads
\beq\label{WKBformula}
	\frac{\imo Q_{0}}{\sqrt{2 Q_{0}''}}
	- \sum_{k=2}^{6}\Lambda_{k} = n+\frac{1}{2},\qquad n=0,1,2\ldots,
\eeq
  where the correction terms $\Lambda_{k}$ were obtained in \cite{WKB,
  WKBorder}.  Here $Q_{0}$ and $Q_{0}^{k}$ are the value and the $k$-th
  derivative of $Q$ at its maximum with respect to the tortoise coordinate
  $r_*$, and $n$ labels the overtones. The WKB formula (\ref{WKBformula})
  was effectively used in a lot of papers (see, e.g., \cite{WKBuse} and
  references therein).

\begin{table}
\caption{Quasinormal modes of axial perturbations of black universes with
 	$c=-1$ ($m=1/3$).}
\begin{tabular}{|c|c|c|}
\hline
$\ell$&WKB6&time-domain fit\\
\hline
$2$&&$0.712299-0.157848\imo$\\
$3$&$1.168844-0.176095\imo$&$1.169144-0.176019\imo$\\
$4$&$1.590954-0.182539\imo$&$1.590956-0.182542\imo$\\
$5$&$1.997792-0.185591\imo$&$1.997801-0.185588\imo$\\
$\gg1$&$0.191226(2\ell+1-\imo)$&\\
\hline
\end{tabular}\label{tabl}
\end{table}

  The WKB formula gives an accurate result for large multipole numbers (see
  Table \ref{tabl}). Expanding the WKB formula it powers of $\ell$, we find
  the following asymptotical expression for $c=-1$:
\[
  \omega b = \sqrt{\frac{2}{\pi}\tan^{-1}\left(\frac{2}{\pi}\right)}
	\left(\ell+\frac{1}{2}-\imo\Big(n+\frac{1}{2}\Big)\right)
	+ {\cal O}\left(\frac{1}{\ell}\right).
\]
  In Table \ref{tabl}, the asymptotic formula for the fundamental mode
  (the one that dominates at late times) is presented in units $m=1/3$,
  so that $b=\pi/2$.

  The WKB formula (\ref{WKBformula}) used here is developed for effective
  potentials which have the form of a barrier with only one peak (see, e.g.,
  Fig.\,\ref{_fig4} for black universes), so that there are two turning
  points given by the equation $\omega^2 - V_{eff} =0$. Therefore, it cannot
  be used for effective potentials which have negative gaps, that is, for
  testing the stability of all questionable cases.

\section{Conclusions}

  We have developed a general formalism for analyzing axial gravitational
  perturbations of an arbitrary \ssph\ solution to the
  Einstein-Maxwell-scalar equations where the scalar field, which can be
  both normal and phantom, is minimally coupled to gravity and possesses
  an arbitrary potential. This can be used for studying the stability and
  QNM modes of diverse charged and neutral scalar field configurations in
  \GR. As to polar perturbations, we have restricted ourselves to the
  monopole mode, i.e., to \sph\ (radial, for short) perturbations.

  We have applied this formalism to some electrically neutral \whs\ and \bhs\
  supported by a phantom scalar field. The main results which were obtained
  here are:
\begin{enumerate}
\item
    M-AdS wormholes described by the solution (\ref{sol-pha1}) with negative
    mass are shown to be unstable under radial perturbations, although the
    initial effective potential with a singularity at the throat is positive
    everywhere. These features are similar to those of anti-Fisher \whs\
    \cite{Gonzalez:2008wd} which are twice \asflat.
\item
    Black universes (i.e., regular \bhs\ with \dS\ expansion far beyond
    the horizon), described by the solution (\ref{sol-pha1}) without throats
    in the static region, are shown to be stable only in the special case
    where the horizon coincides with the minimum of the area function $R(r)$
    (the parameters: $c=-1,\ b=3\pi m/2$) and unstable for $c\ne -1$.
\item
    Quasinormal modes of axial perturbations have been calculated for
    stable black universes.
\end{enumerate}

  Quite a lot of other problems of interest are yet to be studied. One can
  mention a full nonlinear analysis of perturbations in all relevant cases.
  Next, the formalism developed here for linear perturbations allows us to
  include the electromagnetic field into consideration and to study charged
  solutions with both normal and phantom scalar fields. Last but not least,
  using the well-known conformal mappings that relate Einstein and Jordan
  frames of scalar-tensor and curvature-nonlinear theories of gravity, one
  can extend the studies to solutions of these theories.

\begin{acknowledgments}
 The work of K.B. was supported in part by RFBR grant 09-02-00677a, by NPK
 MU grant at PFUR, and by FTsP ``Nauchnye i nauchno-pedagogicheskie kadry
 innovatsionnoy Rossii'' for the years 2009-2013.  R. K. acknowledges the
 Centro de Estudios Cientifcos (Valdivia, Chile) where part of this work was
 done. At its initial stage the work was partially funded by the Conicyt
 grant ACT-91: ``Southern Theoretical Physics Laboratory'' (STPLab). The
 Centro de Estudios Cientifcos (CECS) is funded by the Chilean Government
 through the Centers of Excellence Base Financing Program of Conicyt. At the
 final stage R. K. was supported by the European Commission grant through
 the Marie Curie International Incoming Contract.  A. Z. was supported by
 Deutscher Academischer Austausch Dienst (DAAD) and Conselho Nacional de
 Desenvolvimento Cient\'ifico e Tecnol\'ogico (CNPq).
\end{acknowledgments}

\end{document}